\documentclass[apj]{emulateapj}
\usepackage{graphicx}
\usepackage{natbib}
\usepackage{times}
\usepackage[normalem]{ulem}
\usepackage[colorlinks=true,urlcolor=blue,citecolor=blue,linkcolor=blue]{hyperref}
\usepackage{aas_macros}
\usepackage{microtype}
\newcommand{\ha}{H$\alpha$}

\newcommand{\calong}{\ion{Ca}{2}~8542~\AA}
\newcommand{\mglong}{\ion{Mg}{2}~2976~\AA} 
\newcommand{\silong}{\ion{Si}{4}~1403~\AA}

\newcommand{\ca}{\ion{Ca}{2}}
 
\newcommand{\si}{\ion{Si}{4}}

\newcommand{\iris}{\textit{IRIS}}
\newcommand{\kms}{km~s$^{-1}$}
\newcommand{\xt}{$xt$}
\newcommand{\lt}{$\lambda t$}
\begin{document}

\title{On the active region bright grains observed in the transition region imaging channels of \iris}
\author{H. Skogsrud\altaffilmark{1}}
\author{L. Rouppe van der Voort\altaffilmark{1}}
\author{B. De Pontieu\altaffilmark{2,1}}
\affil{\altaffilmark{1}Institute of Theoretical Astrophysics, University of Oslo, P.O. Box 1029 Blindern, N-0315 Oslo, Norway}
\affil{\altaffilmark{2}Lockheed Martin Solar \& Astrophysics Lab, Org.\ A021S, Bldg.\ 252, 3251 Hanover Street Palo Alto, CA~94304 USA}

\begin{abstract}The Interface Region Imaging Spectrograph (\iris) provides spectroscopy and narrow band slit-jaw (SJI) imaging of the solar chromosphere and transition region at unprecedented spatial and temporal resolution. Combined with high-resolution context spectral imaging of the photosphere and chromosphere as provided by the Swedish 1--m Solar Telescope (SST), we can now effectively trace dynamic phenomena through large parts of the solar atmosphere in both space and time. \iris\ SJI 1400 images from active regions, which primarily sample the transition region with the \ion{Si}{4} 1394 and 1403~\AA\ lines, reveal ubiquitous bright ``grains'' which are short-lived (2--5 min) bright roundish small patches of sizes 0.5-1.7\arcsec\ that generally move limbward with velocities up to about 30~\kms. In this paper we show that many bright grains are the result of chromospheric shocks impacting the transition region. These shocks are associated with dynamic fibrils (DFs), most commonly observed in \ha. We find that the grains show strongest emission in the ascending phase of the DF, that the emission is strongest towards the top of the DF and that the grains correspond to a blueshift and broadening of the \ion{Si}{4} lines. We note that the SJI 1400 grains can also be observed in the SJI 1330 channel which is dominated by \ion{C}{2} lines. Our observations show that a significant part of the active region transition region dynamics is driven from the chromosphere below rather than from coronal activity above. We conclude that the shocks that drive DFs also play an important role in the heating of the upper chromosphere and lower transition region.
\end{abstract}

\section{Introduction}
\label{Sec:intro}

The active region chromosphere observed in \ha\ is dominated by fibrilar structures of various lengths. While the longest fibrils are static and low-lying structures that appear to connect opposite magnetic polarity plage regions \citep[e.g., ][]{van_noort_2006, de-pontieu_et_al_2007c, guglielmino_et_al_2010, kuridze2011, 2011ApJ...742..119R}, the shorter fibrils (1--4~Mm) are short lived (3--6~min) and display a characteristic up and down motion. These shorter fibrils are commonly referred to as dynamic fibrils (DFs). They often recur (semi-periodically) and display little sideways motion so that their dynamical evolution can be well measured along a linear trace through the spatio-temporal domain by means of space-time diagrams.

\cite{hansteen_et_al_2006} and \cite{de-pontieu_et_al_2007a} showed that in \xt--diagrams, the top of the DFs follows a parabolic trajectory with a deceleration that is typically only a fraction of solar gravity. They used advanced numerical radiative magneto-hydrodynamical simulations to demonstrate that DFs are formed by chromospheric shocks that occur when waves caused by convective flows or global oscillations leak into the chromosphere along inclined magnetic field lines as previously suggested by \cite{michalitsanos_1973, bel_et_al_1977, suematsu_et_al_1990} and \cite{de-pontieu_et_al_2004, de-pontieu_et_al_2005}. 

\cite{heggland_et_al_2007, heggland_et_al_2011} studied shockwave driven jets in more detail and found that correlations between velocity and deceleration of the jets in their numerical simulations matched well with DF observations which support the view that DFs are driven by magneto-acoustic shocks. \citet{martinez-sykora_et_al_2009} studied spicules in a 3D numerical simulation that covered the solar atmosphere from the upper layers of the convection zone to the lower corona. These simulated spicules described parabolic trajectories similar to DFs and were found to be driven by a variety of mechanisms that include $p$-modes, collapsing granules, magnetic energy release in the photosphere and lower chromosphere, and convective buffeting of flux concentrations. 

Parabolic trajectories and correlations between velocity and deceleration similar to DFs were observed for: some mottles in quiet Sun network \citep{rouppe_van_der_voort_et_al_2007}, a subset of spicules at the limb \citep[so-called type \footnotesize I\normalsize spicules, ][]{de-pontieu_et_al_2007b,tiago_et_al_2012}, in small-scale jets in the chromospheres of sunspots \citep{rouppe_van_der_voort_et_al_2013,yurchyshyn_et_al_2014,2014ApJ...789..108C}, in sunspot oscillations observed with IRIS \citep{2014ApJ...786..137T} and in spicules observed above a plage region with Hinode \citep{2010PASJ...62..871A}. \cite{2009A&A...499..917K} studied DFs in Ly $\alpha$ where they appear bright against a dark background, which is in contrast to \ha\ where they appear dark against a bright background.

\cite{langangen_et_al_2008_2, langangen_et_al_2008_1} studied DFs spectroscopically and identified their characteristic shock signature in the spectral evolution (\lt--diagrams): a sudden (strong) blueshift of the line, followed by a gradual (linear) shift of the line to redshifts of typically $+15$\kms. The transverse motion of DFs in \ha\ is studied by \cite{2007ASPC..368..115K} and a rebound shock mechanism for fibrils is presented by \cite{1989ApJ...343..985S}.

There are few studies which analyses the thermal evolution of DFs. Some small scale structures in images from transition region and coronal channels have been studied in the literature, but they are either unrelated features \citep{2013ApJ...770L...1T,2014ApJ...784..134R,2014ApJ...789L..42K,2014ApJ...790L..29T,2014Sci...346A.315T}, using EUV observations with insufficient spatio-temporal resolution to connect the features to DFs \citep{2000ApJ...535L..63W,2011A&A...532L...1M} or lacking co-observation in chromospheric diagnostics passbands \citep{2015ApJ...807...71P,2015ApJ...809L..17J}. \cite{de_wijn_et_al_2006} studied DFs in \ha\ and the 1550~\AA\ filter from the Transition Region and Coronal Explorer \citep[TRACE, ][]{handy_et_al_1999} which contains contributions from the transition region line \ion{C}{4}. The authors found a correlation between the DFs in \ha\ and 1550 passband emission that is best in the downward falling phase. This correlation suggests that some plasma associated with the DFs is heated to transition region temperatures. Further characterization of the role of DFs in upper atmospheric heating has been difficult due to the observational challenges of resolving the fundamental temporal and spatial time scales of the chromosphere, transition region and corona. A major breakthrough in our observational capabilities is currently provided by the Interface Region Imaging Spectrograph \citep[IRIS, ][]{de-pontieu_et_al_2014_iris}, launched in July 2013. \iris\ provides spectroscopy and narrow band slit-jaw imaging of the chromosphere and transition region at unprecedented spatial and temporal resolution. Combined with high-resolution context spectral imaging of the photosphere and chromosphere as provided by the Swedish 1-m Solar Telescope \citep[SST, ][]{scharmer_et_al_2003} on La Palma, we can now effectively trace the upper atmospheric response to DFs.


\section{The Observations}
\label{sec:obs}

Three \iris\ datasets are used in this study, details are provided in table~\ref{tab:datasets}. All of the datasets target active regions containing plage and all have co-observations from the Swedish 1-m Solar Telescope \citep[SST, ][]{scharmer_et_al_2003}. Two of the \iris\ datasets are 4-step dense rasters with a cadence of 21~s, the spatial step width is 0\farcs33, equivalent to the spatial width of the spectrograph slit. Slit-jaw images were recorded at the same cadence as the spectrograph rasters. The third dataset is a large dense spatial raster with 96 slit positions covering a region of 31\arcsec $\times$ 175\arcsec\ in size. The pixel scale of the slit-jaw images are 0\farcs17~pix$^{-1}$. Different spectral windows were covered by the spectrograph, of which the windows that cover \ion{Si}{4}~1403~\AA\ (spectral pixel scale 12.8~m\AA) and the \ion{Mg}{2}~h\&k lines (spectral pixel scale 25.6~m\AA) are of prime interest for our analysis. The SJI images recorded are from the SJI 2796, SJI 2832 and SJI 1400 channels, all with 4~s exposure time. SJI 2796 covers the \ion{Mg}{2}~k line core (FWHM 4~\AA, chromosphere), SJI 2832 the wing of the \ion{Mg}{2} h line (FWHM 4~\AA, photosphere) and SJI 1400 the \ion{Si}{4} 1394 and 1403~\AA\ spectral lines (FWHM 55~\AA, transition region). These channels are not pure but contain a mix of several contributions including continuum (for the 1400/1330 channels) and photospheric inner wings (for the 2796 channel). The June 11, 2014 dataset also includes observations in the SJI 1330 channel, which covers the \ion{C}{2} lines (FWHM 55~\AA, upper-chromosphere/transition region).

We analyze level 2 \iris\ data, the standard data product that is publicly available in the \iris\ database and for which the calibration includes dark, flat field and geometrical corrections. We improved on the standard wavelength calibration of the spectra: for the June 11 dataset we shifted the spectra by 40.8~m\AA\ so that the optically thin \ion{O}{1}~1356~\AA\ line is at its vacuum laboratory wavelength of 1355.5977~\AA. None of the observed spectral windows of the September 15 dataset covered the standard reference calibration lines. We therefore shifted the \silong\ spectra by 71.6~m\AA\ so that the average \silong\ Doppler shift of all the spectra matched the June 11 average shift. This leaves uncertainty in the wavelength calibration, but we note that we are not studying in detail the absolute Doppler shift of the \silong\ line, only the excursions from the mean line profile. We do not use the spectrograph data of the September 9 dataset and consequently we kept the superior spatial resolution of the SST (0\farcs058~pix$^{-1}$) and interpolated the \iris\ data to the SST pixel scale. For more information on \iris\ and its instruments, we refer to \citet{de-pontieu_et_al_2014_iris}

The SST data consist of observations in the \calong\ and \ha\ spectral lines obtained with the Crisp Imaging SpectroPolarimeter \citep[CRISP, ][]{scharmer_2006}. \calong\ was sampled at 25 line positions (filter FWHM 111 m\AA), from $-1200$~m\AA\ to $+1200$~m\AA\ ($\pm 42$~\kms, relative to the line core) in steps of 100~m\AA, and \ha\ was sampled at 15 line positions (filter FWHM 66~m\AA) from $-1400$~m\AA\ to $+1400$~m\AA\ ($\pm 64$~\kms) in steps of 200~m\AA. In addition, maps of the 4 Stokes parameters were acquired at $-48$~m\AA\ in the blue wing of the \ion{Fe}{1}~6302 line. The time to complete the spectral scan of all lines was on average 11~s. 

The high spatial resolution in the SST data (0\farcs058~pix$^{-1}$) was achieved with aid of the adaptive optics system \citep{scharmer_et_al_2003a} and image restoration with Multi-Object Multi-Frame Blind Deconvolution \cite[MOMFBD, ][]{van_noort_et_al_2005}. We followed the CRISPRED \citep{crispred} data reduction pipeline to process the data. The MOMFBD restoration of a spectral line scan included 8 exposures per spectral line position for the two CRISP cameras and for the wide band camera that is positioned before the CRISP Fabry-P\'erot instrument but after the CRISP pre-filter (FWHM 0.5~\AA\ for \ha\ and 1~\AA\ for \ca, note that the wide band camera has synchronized exposures with the CRISP cameras). The inclusion of the wide band channel in the MOMFBD restoration ensures accurate alignment between the sequentially recorded spectral line positions and facilitates a high degree of spectral integrity \citep[also see, ][]{henriques_2012}.

We further used data from the Atmospheric Imaging Assembly \citep[AIA, ][]{lemen_et_al_2012} on board the Solar Dynamics Observatory \citep[\textit{SDO}, ][]{pesnell2012} for contextual reference. We use the 1600~\AA\ channel for alignment with the \iris\ and SST and the 304~\AA, 171~\AA, 193~\AA, 211~\AA\ and 94~\AA\ channels. 

For \iris/SST alignment of the September data, we used the \ha\ $+1400$~m\AA\ wing position and the photospheric \iris\ SJI 2832 as reference, after scaling down the SST data to the \iris\ pixel size, 0\farcs17~pix$^{-1}$. For the June 11 dataset, we aligned the SST \calong\ $-1200$~m\AA\ wing position to the \ion{Mg}{2}~k \iris\ SJI 2796. For the alignment of the AIA data, we used the AIA 1600 channel as reference (through cross-correlation with \iris\ SJI 2796). We find the accuracy of the alignment to be on the order of one \iris\ pixel.

For the analysis of the final aligned data we made extensive use of CRISPEX \citep{vissers_et_al_2012}, a graphical tool for effective browsing through multi-dimensional data. CRISPEX is available in SolarSoft under the \iris\ tree. 

\begin{table*}[bth]
\caption{Overview of the \iris\ datasets analyzed in this study.}
\begin{center}
\begin{tabular}{cccccccc}
	\hline \hline
& & & & & \multicolumn{3}{c}{overlap with SST$^{\rm{c}}$} \\
 Date & Time (UT) & Type$^{\rm{a}}$ & FOV$^{\rm{b}}$ & Pointing & Time & SJI overlap $^{\rm{d}}$ & Slit \\
 	\hline
11-June-2014 & 07:36 -- 10:28 & Dense synoptic 96-step raster & 31\arcsec $\times$ 175\arcsec & AR 12080 (574\arcsec, -200\arcsec) & 02:52  & 100\% & 33\arcsec-63\arcsec \\  
9-Sep-2014   & 07:59 -- 10:59 & Medium dense 4-step raster & 1\arcsec $\times$ 60\arcsec & AR 12157 (-230\arcsec, -367\arcsec) & 01:50  & 93\% & 55\arcsec \\
 15-Sep-2014 & 07:49 -- 10:59 & Medium dense 4-step raster & 1\arcsec $\times$ 60\arcsec  & AR 12158 (743\arcsec, 163\arcsec)  & 01:16 & 80\%  & 56\arcsec \\
   	\hline
\end{tabular}
\begin{minipage}{.9\hsize}
  $^{\rm{a}}$ Type of observation. \\
  $^{\rm{b}}$ Area covered by the spectrograph slit. \\
  $^{\rm{c}}$ The last three columns provide information on the overlap of the SST and \iris\ data. \\
  $^{\rm{d}}$ Percentage of SST pixels covered in the \iris\ SJI FOV.
\end{minipage}
\end{center}
\label{tab:datasets}
\end{table*}

\section{Results}
\label{sec:results}

\begin{figure*}
  \centering
  \includegraphics[width=0.8\textwidth]{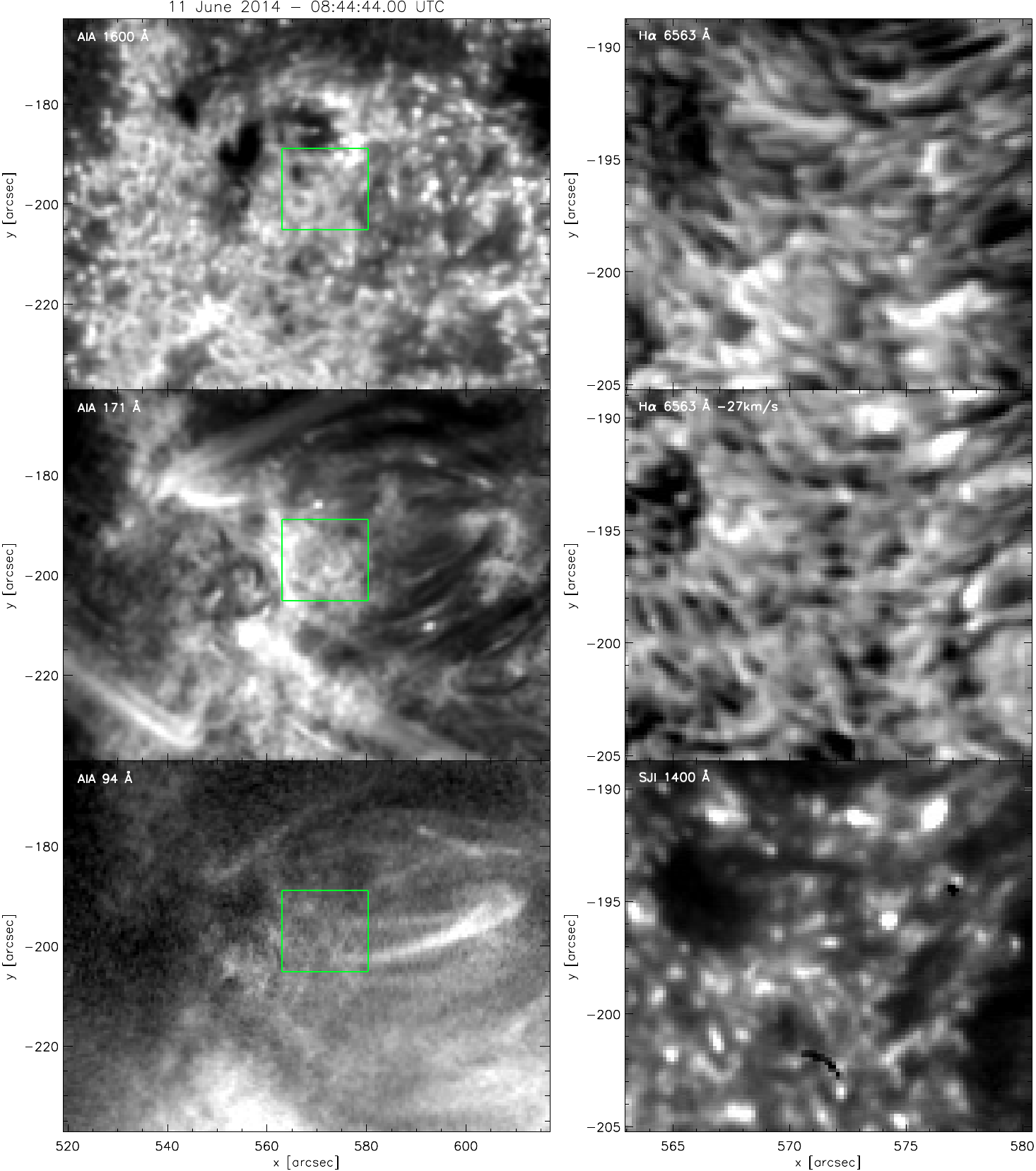}
  \caption{IRIS SJI 1400 grains in active region plage with DFs. The left column shows AIA overview images of AR12080 on June 11, 2014, from top to bottom: AIA~1600~\AA, AIA~171~\AA, and AIA~94~\AA. These AIA images are shown with square root intensity scaling to give more weight to weaker background emission in the plage region. The central square marks the region filled with DFs shown at larger magnification at right, from top to bottom: \ha\ line core, \ha\ $-600$~m\AA\, and SJI 1400~\AA. An animation of this figure, as well as an animation alternating between \ha\ line core and the inverse of SJI 1400~\AA, are available in the online material.}
  \label{fig:fov}
\end{figure*}

In all \iris\ SJI 1400 observations of active region plage with DFs, the field of view is dominated by small-scale bright grains. The grains are typically between 0.5--1.7\arcsec\ in size with most being smaller than 1\arcsec. Their lifetimes are typically between 2--5~minutes with most having a lifetime around 3~minutes. While there is some erratic motion associated with these grains, the general impression is that of a motion in the limb-ward direction. We note that similar grains are present in the SJI 1330 channel which is tightly correlated to the SJI 1400 grains. In the remainder we concentrate on the SJI 1400 channel for which all our datasets have associated \ion{Si}{4} spectra. The morphology and temporal behavior of the SJI 1400 grains are illustrated in Fig.~\ref{fig:fov} and associated animations.

Fig.~\ref{fig:fov} shows an overview of the field of view from the June 11 dataset. The observation targets a region containing plage close to several sunspots. All of the datasets are ``mossy'' in part \citep{berger_et_al_1999}; the AIA 171 and AIA~94 images show that both hot and cool coronal loops have foot points in the locations where we find DFs and we find a bright, reticulated pattern in the AIA~171 channel. In the column to the right we zoom in on a smaller field of view in the \ha\ and SJI 1400 passbands. The dark elongated patches in the \ha\ images are DFs and the brightest roundish ``spots'' in the SJI 1400 image are the bright grains. In the associated animation online, we show the AIA~304 channel instead of the AIA~94 channel. The characteristic up and down motion of the DFs is clearly visible in \ha\ line core while in the AIA~171 \& 304 panels we can see indications of small scale motion which could be related to DFs. We do not see any similar motion in the AIA~1600 movie, see the Discussion section. In the blue wing one mostly sees the upward motion of the DFs -- they disappear once they reach maximum extension. Careful visual comparison between the \ha\ panels and the SJI 1400~\AA\ panel reveals that many SJI 1400 grains are associated with DFs -- but not all \ha\ DFs have associated SJI 1400 signal. A second animation associated with Fig.~\ref{fig:fov} shows the time series in \ha\ line core and SJI 1400 with inverted color scheme -- this makes it easier to make a visual connection between the dark DFs in \ha\ line core and SJI 1400 grains as dark patches. The third panel in the animation alternates between the two diagnostics which highlights the similar dynamical properties between the channels.

\begin{figure}
  \centering
  \includegraphics[width=\columnwidth]{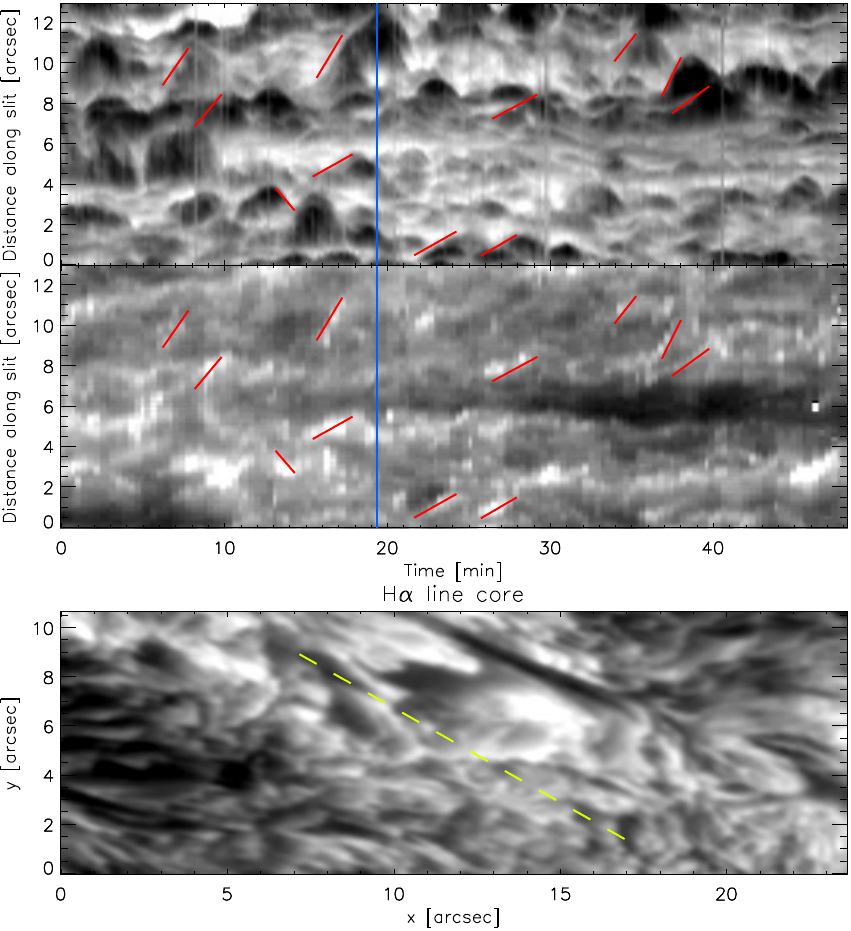}
  \caption{ An \xt--diagram with a long time range along a fixed 1\arcsec\ wide artificial slit in the September 9 dataset. The slit is oriented in the general direction of the primary axis of the DFs (see associated animation) and is shown as a green dashed line in the x--y snapshot in the bottom panel. The snapshot is from the time marked with blue vertical lines in the other panels. The top panel shows the \ha\ line core and the middle panel shows \silong\ with a gamma factor of 0.2. The red lines, which have the same positions in both panels, are visual aids to illustrate the tight correspondence of the dark \ha\ parabolas with the bright ridges in \si. Note that the red lines show only a sub selection of the features with good correspondence. The vertical gray stripes in the \ha\ panel are due to moments with bad seeing. Two animations of this figure are available in the online material.}
  \label{fig:long-xt}
\end{figure}

The tight correspondence between DFs and bright grains is further illustrated in Fig.~\ref{fig:long-xt} where we show an \xt--diagram with a long time range from a fixed artificial slit 1\arcsec\ wide. The average value across the slit was computed per time step and is shown in the figure. For this figure we kept the superior spatial and temporal cadence of SST and scaled up the \iris\ SJI 1400 to the SST image scale of 0\farcs058. The top panel shows dark parabolas which are the trademark signature of the DFs in \ha\ and in the bottom panel we show the corresponding SJI 1400 \xt--diagram where we have applied a gamma factor of 0.2 to the intensity ($I^{0.2}$) -- this softens the sharp contrast in this channel. When looking at the SJI movies the bright grains are relatively bright and the dominant feature in plage. The red lines trace out the initial \ha\ phase and serves as visual aid. The bright ridges in SJI 1400 match the ascent phase of the DF well and usually appear toward the top of the DF; the later phase also match well between the two channels, however the SJI 1400 emission is significantly weaker compared to the initial phase. The tight correlation between the \ha\ and SJI 1400 emission is particularly clear from viewing the associated animation which blinks the two diagrams against each other. The dark band in the SJI 1400 panel at 7\arcsec\ from about 30~min and onward is caused by a small dark patch in the SJI image almost devoid of emission; similar to the region at coordinates (565\arcsec,-195\arcsec) in the SJI 1400 panel of Fig.~\ref{fig:fov}.

\begin{figure} 
  \centering
  \includegraphics[width=\columnwidth]{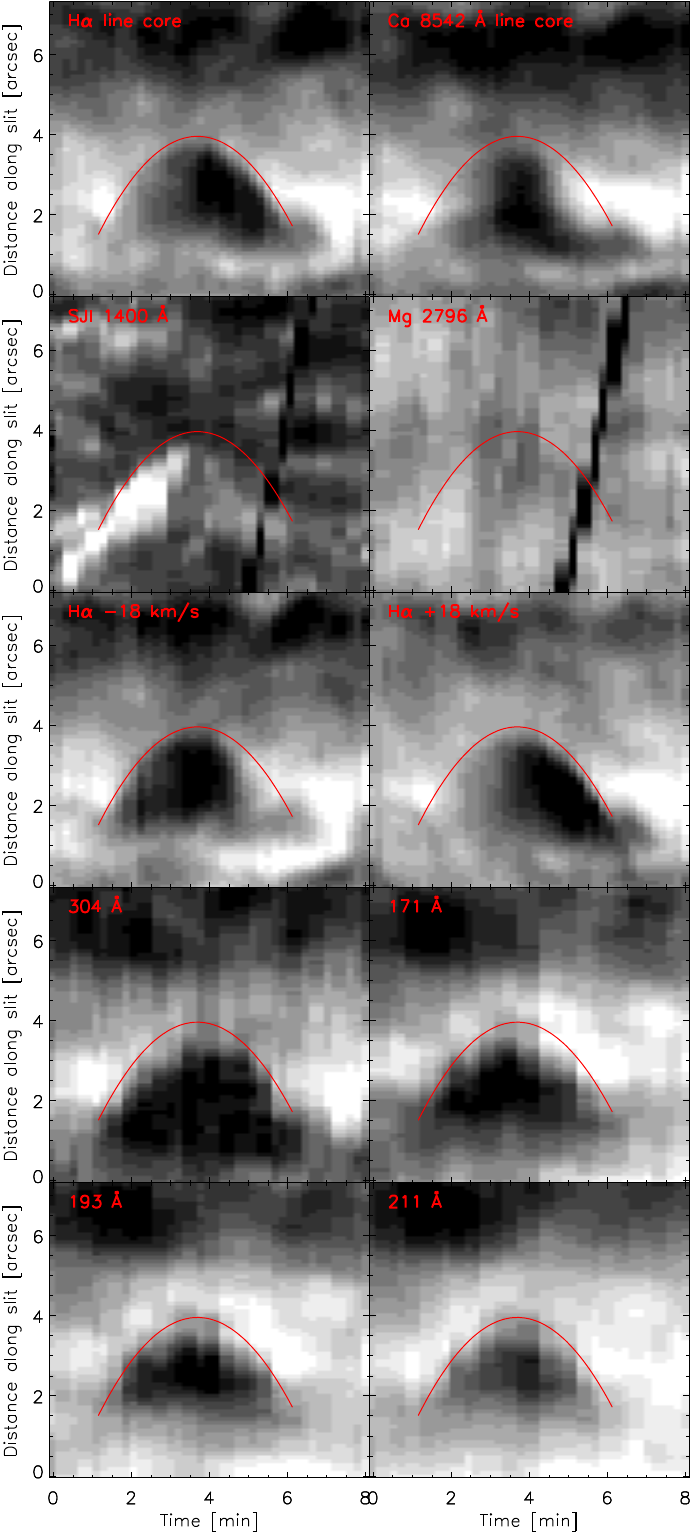}
  \caption{An \xt--diagram showing the behavior of one DF in multiple passbands along a 1\arcsec\ wide artificial slit from the June 11 dataset. The red parabola traces out the \ha\ parabola. The dark, inclined line in the \silong\ and \mglong\ passbands is caused by the spectrograph slit. The artificial slit is fixed in space. An animation of this figure, including x--y snapshots, is available in the online material.}
  \label{fig:xt}
\end{figure}

In Fig.~\ref{fig:xt} we show \xt--diagrams for one DF in several passbands, including four AIA channels. For the construction of the \xt--diagrams we used a 1\arcsec\ wide artificial slit from which we show the mean value. The \ha\ DF is traced out by the red parabola which is overplotted in all panels. The DF is clearly visible in all passbands except in \mglong. The SJI 1400 emission is strongest during the initial ascent phase of the parabola, where also the 304~\AA\ channel from AIA shows stronger emission. The \ha\ blue and red wing images clearly illustrate the transition from blue shift in the initial phase to redshift in the later stages. In the AIA channels the red parabolas trace out dark absorbing features caused by bound-free hydrogen and helium continuum absorption.

\begin{figure*}
  \centering
  \includegraphics[width=0.85\textwidth]{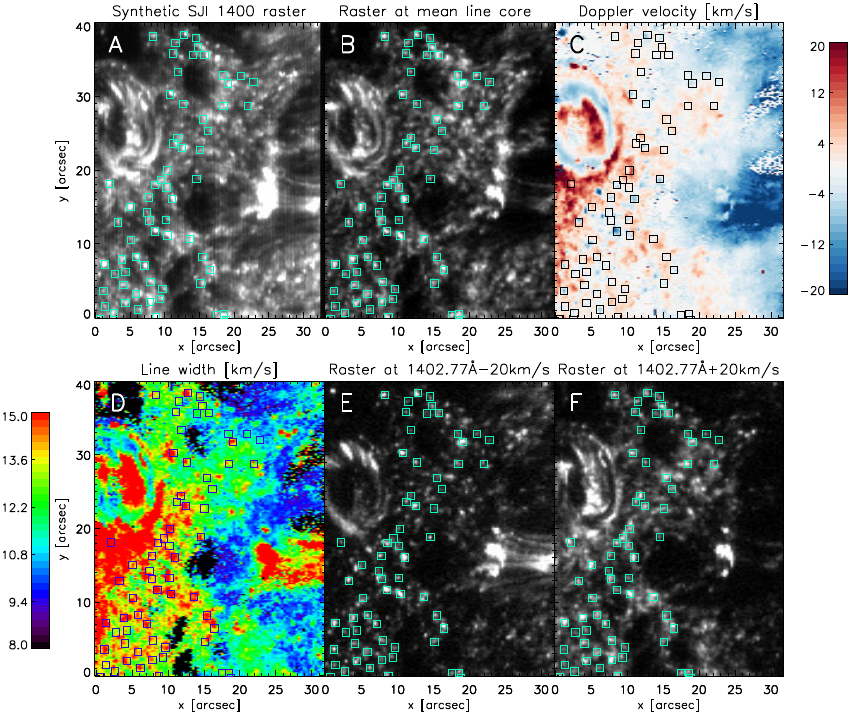}
  \caption{\ion{Si}{4}~1403~\AA\ raster images from the June 11 dataset. \textit{(Top row)} Panel A shows a synthetic SJI 1400 raster constructed by extracting rectangular slices around the slit from the SJI images closest in time (see text for full explanation). Each column can be directly compared to the other rasters constructed from spectrograph data because they are co-spatial and nearly co-temporal. Panel B shows the intensity at the average wavelength position of the \ion{Si}{4} 1403~\AA\ line core. Panel C shows the computed Doppler velocity with respect to this average wavelength position. \textit{(Bottom row)} From left to right, the panels show the line width of the \silong\ line and two rasters from $\pm 20$~\kms\ around 1402.77~\AA, the vacuum laboratory wavelength of the \silong\ line. The small squares mark the grains found with an automated detection algorithm.}
  \label{fig:rasters}
\end{figure*}

In Fig.~\ref{fig:rasters} we turn our attention to the spectral information in our data. To determine whether the grains are due to \ion{Si}{4} 1394/1403 \AA\ line emission or continuum emission we constructed a synthetic raster from the SJI 1400 images, shown in panel A. For each slit position of the spectral raster we located the SJI image closest in time and extracted a 0\farcs33 (2 pixel wide column) vertical slice which we placed successively from left to right in the synthetic raster. Since the SJI images were taken at slower temporal cadence than the spectra, multiple vertical slices were taken from the same SJI image. This includes regions just adjacent to the spectrograph slit in the images. A little amount of the light entering the \iris\ spectrograph is reflected to the SJI images and this weak signal at the location of the spectrograph slit in the slit jaw images was enhanced by division by an average slit profile. This correction resulted in a rather even intensity distribution in the constructed raster although some vertical stripes are visible as artifacts. Each column can be directly compared to the other rasters constructed from spectrograph data because they are co-spatial and nearly co-temporal. 

In panel B of Fig.~\ref{fig:rasters}, we show the intensity at the average line position of the \silong\ line. The large degree of similarity with the slit jaw raster, both in morphology and to large extent in contrast, demonstrates that most signal in SJI 1400 in this region stems from \ion{Si}{4} line emission rather than continuum contribution. We show the Doppler shift and line width computed by fitting a single peaked Gaussian function to the line profile; $I(\lambda)=k + h\exp^{-0.5(\lambda-\mu)^2/\sigma^2}$, where $h$ is the intensity, $k$ is the background signal, $\mu$ is the line center wavelength and $\sigma$ is the line width. We set a lower intensity threshold to avoid computing the parameters on profiles with very little signal. The threshold was set to 20\% of the integrated average \silong\ spectrum and compared to each integrated single profile. 

\begin{figure*}
  \centering
  \includegraphics[width=0.8\textwidth]{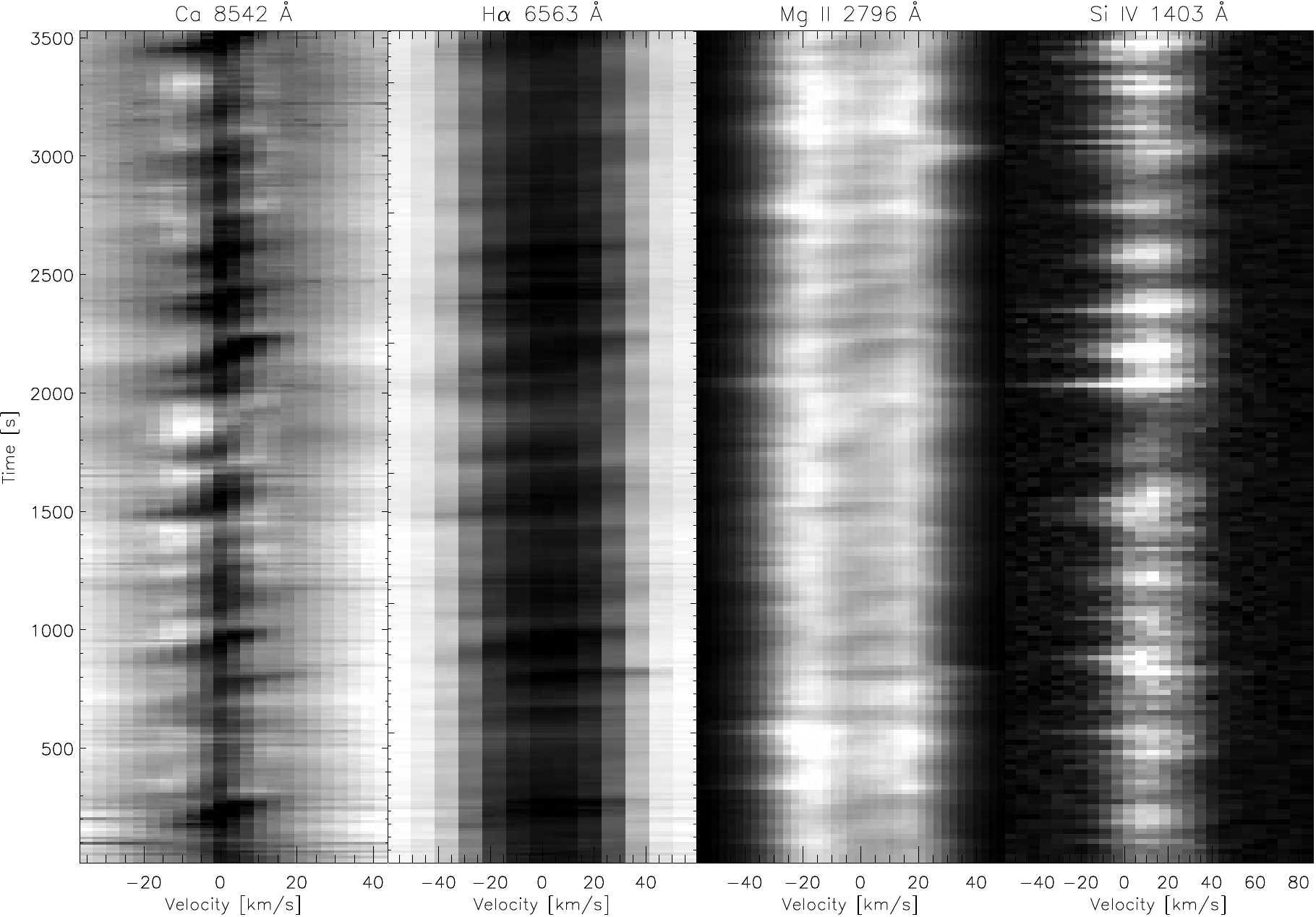}
  \caption{Spectral evolution at a spatial location with DFs in the September 15 dataset. From left to right: \lt--diagrams with long time range in \calong, \ha, \ion{Mg}{2}~k, and \silong. The characteristic sawtooth shock pattern of the DFs is clearly seen in the \calong\ time-series.}
  \label{fig:lt}
\end{figure*}

Visual inspection of panel C reveals that many grains, outlined by small squares, coincide with blueshifted patches. To determine whether there is a systematic trend in the Doppler velocity and spectral width of the bright grains compared with the surroundings, we compared the histograms of both quantities for the two different regions. We only did this analysis on the parts of the FOV which shows short dynamic fibrils in \ha. This resulted in the exclusion of the blue region to the right of about $x=15$\arcsec\ in the upper right panel and the small loops centered on $[x,y]=[5\arcsec, 25\arcsec]$. To automatically detect the bright grains, 
we enhanced the raster image in panel B with the \`a trous algorithm described in \cite{starck}, thresholded the enhanced image at the percentile value of 97\% and removed detections containing two or less pixels. This resulted in the detection of 86 bright grains; the central location of the grains is marked with small squares in all panels. For all the pixels belonging to a single grain we extracted the minimum signed Doppler velocity (i.e., most blueward velocity) and the maximum spectral width. We compared the velocity histogram of the grains to the histogram of all other pixels and find a statistically significant average blueward shift of 3.7~\kms, from 4.9 to 1.2~\kms. For the spectral widths, we find a statistically significant enhancement of the average width by 3.7~\kms, from 14.0 to 17.7~\kms. We verified that if we instead measure the mean grain velocity and width (instead of min/max) and compare to all other pixels that the resulting difference between the means of the distributions is statistically significant. We do not expect a perfect correlation between grain intensity and blueshift since the grains observed in the rasters are observed in differents phases of their evolution, and only the very early phase show strong blueshifts (also see the \lt--diagrams in Fig.~\ref{fig:lt}).  

In panels E and F of Fig. ~\ref{fig:rasters} we show rasters from $\pm 20$~\kms\ around nominal line center at 1402.77~\AA\ to illustrate the asymmetry between the wings. The blue-wing wing raster shows primarily only the grains and the red wing raster shows both the grains and other features in between the grains. The bright grains are visible in both wings due to the enhanced broadening of the spectral profile. 

In Fig.~\ref{fig:lt} we show the spectral evolution of DFs in \lt--diagrams from the September 15 dataset. The characteristic sawtooth shock-pattern of the DFs is clearly visible in the \calong\ panel. A corresponding co-temporal pattern is also visible in the other spectral lines. In the \ha\ panel the very wide core of the line partly hides the pattern close to line center, but in the wings the pattern is clearly visible. We note that the \silong\ shock signature is primarily only visible during the initial phase and not the entire time-evolution of the DF that is so clearly visible in the other line.

\begin{figure*}
  \centering
  \includegraphics[width=0.85\textwidth]{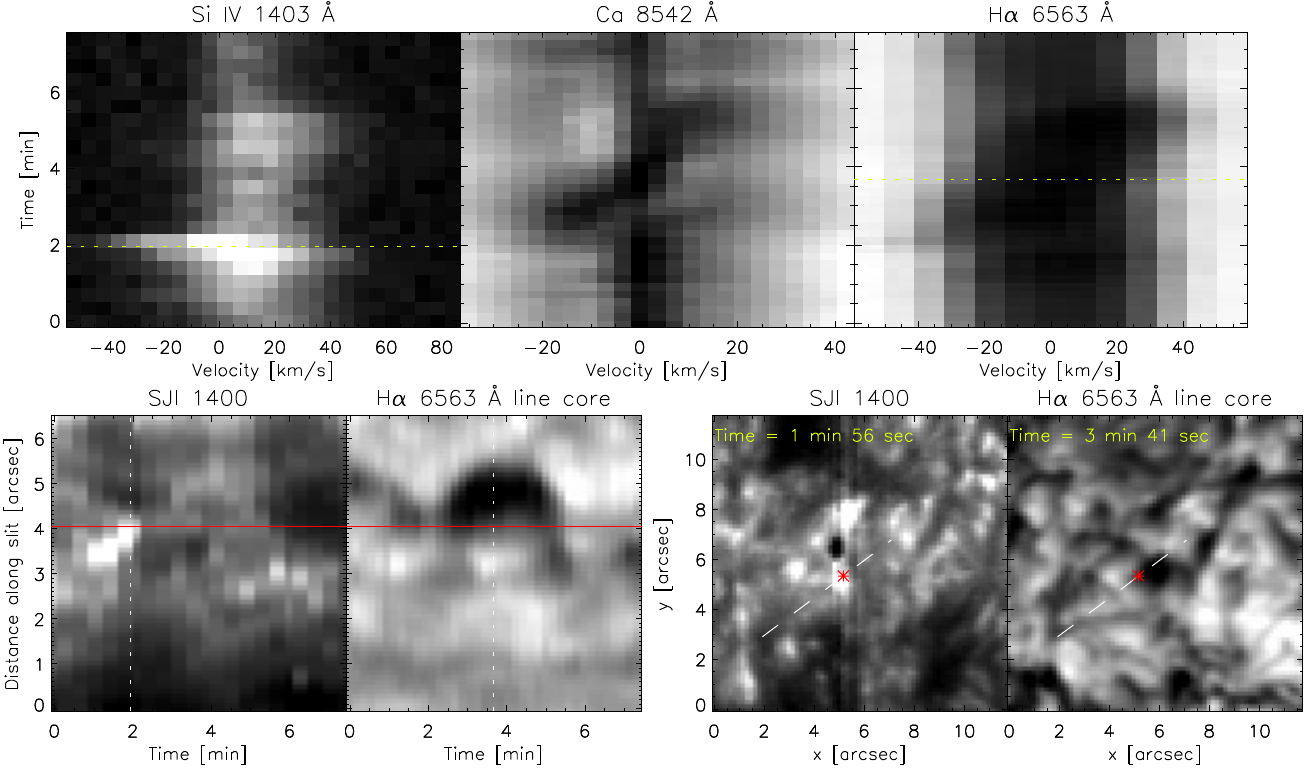}
  \caption{Detailed comparison of the spectral evolution of a single DF in multiple spectral lines. The top row shows \lt--diagrams from the spatial position marked by red asterisks in the lower right x--y snapshots and the red horizontal lines in the bottom left panels. The \xt--diagrams in the lower left show the intensity along the dashed line in the panels to the right and the intensity is the average of a 1\arcsec\ line perpendicular to the line. The x--y snapshots are from the green time step printed in the respective legends and those time positions are marked by green dotted lines in the others panels. An animation of this figure is available in the online material.}
  \label{fig:angle}
\end{figure*}

In Fig.~\ref{fig:angle} we show that the bright \silong\ emission from the grains primarily happens toward the top of the DF. We illustrate this by displaying detailed \lt--diagrams and \xt--diagrams from a DF that crosses the \iris\ slit in the September 15 dataset. In the top row we clearly see the shock signature of the DF in the \calong\ and \ha\ spectral lines as function of time, but in \silong\ we only see a blue shifted brightening in the very early phase of the DF  and a fainter redshift in the end phase of the DF. In the \xt--diagrams in the lower left corner we can see the temporal evolution along the primary axis of the DF and in the lower right corner we see spatial x--y cutouts from two different time steps showing the DF and the bright grain. The SJI 1400 image shows the time when the grain passes the slit; the \lt--diagram at top-left shows that this result in increased \silong\ emission which is blue shifted and broadened. The top of the DF moves toward the upper right corner in the x--y cutouts while the ``body'' of the DF remains covered by the slit during its lifetime resulting in a complete sawtooth pattern in \calong\ and \ha. In \silong\ however, there is little signature of the characteristic deceleration after the initial blueshift $t=2$~min; there is a slight hint of enhanced redshifted \silong\ emission at the final phase of the DF at $t=5$~min. This is because the top of the DF (where most of the SI IV emission originates) no longer is covered by the slit, the Si IV emission only shows the beginning and end phase of the DF.

\begin{figure*}
  \centering
  \includegraphics[width=\textwidth]{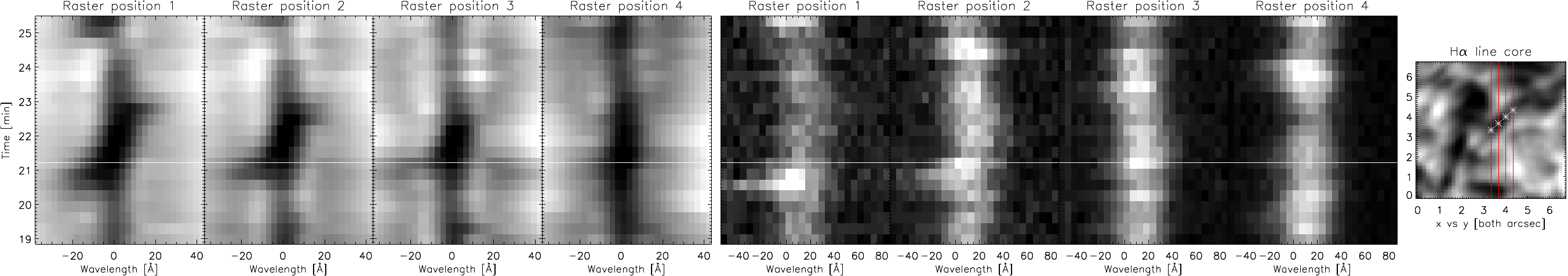}
  \caption{A DF crossing the 4 slit positions in the September 15 dense raster dataset. \lt--diagrams are shown for each slit position at the location marked with asterisks in the rightmost x--y snapshot panel. In the x--y snapshot, the 4 vertical red lines trace out the slit positions in an \ha\ image, which is from the time step marked with a white horizontal line in the other panels. The first 4 panels from left are from the \calong\ line and the next four are from the \silong\ line.}
  \label{fig:4step}
\end{figure*}

In Fig.~\ref{fig:4step} we show the propagation of one DF as it crosses the 4 slit positions in the dense fast rasters from the September 15 data. The spatial location of the slits and the positions of the spectra are marked in the rightmost panel overplotted on a \ha\ line core image. The DF moves from lower left toward the upper right corner and first crosses the leftmost slit (raster position 1). In the \calong\ \lt--diagrams we can clearly see the propagation toward the right because the sharp blueshift associated with the shock appears later for the slit positions to the right. In addition, the phase with characteristic deceleration and increasing redshift is longest for the leftmost slit: this is the slit position that covers the DF for the longest period in its lifetime. The \calong\ spectra indicate that the DF hardly reaches the 4th raster position. In the \silong\ panels the evolution from left to right is also visible: the grain phase is strongest in the left slit: strongest emission, widest profile and strongest asymmetry to the blue. This signature is later and weaker for the 2nd slit, even later and weaker for the 3rd, and virtually absent in the 4th slit. This figure gives another illustration that the \silong\ emission is mostly associated with the top part of the DF: we only see significant \silong\ emission when the top of the DF moves through the subsequent slit positions. While the DF leaves a signature in the \calong\ spectra throughout its lifetime, there is hardly any trace of the DF after the top has passed through in the \silong\ spectra. 

\section{Discussion}
In this paper we use co-observations from the SST and \iris\ that contain photospheric and chromospheric diagnostic spectral lines (\ha, \calong) and transition region lines (\ion{Si}{4}~1394~\AA\ and 1403~\AA) to establish a tight correlation between short-lived active region bright grains observed in the SJI 1400 passband and DFs as seen in \ha\ above plage regions. DFs and bright grains have a tight correspondence in both \lt\ and \xt--diagrams. The grains typically exhibit the strongest emission in the upward ascent phase of the DFs and appear toward the top of the DFs. They generally show emission during the entire lifetime of the DF, although the signal is significantly fainter in the later phase compared to the initial phase. The grain emission in the SJI 1400 channel stems from the \ion{Si}{4} 1394~\AA\ and 1403~\AA\ spectral lines rather than continuum emission. We find that in the grains, the \silong\ spectral profile broadens and is Doppler shifted towards the blue as compared to the surroundings.  \cite{martinez-sykora_et_al_2015} studied a similar set of co-observations of grains observed in internetwork regions and concluded that those grains are due to continuum emission in acoustic shocks, not emission in the \ion{Si}{4} lines.

Whereas DFs can be clearly identified in \ha\ and \calong, in \silong\, we only find clear emission toward the DF top but no distinct spectral signature below along the fibril structure. This means caution should be exercised when analyzing co-observations from close to the limb or where the field direction is more perpendicular to the line of sight. The projection effects may prevent an unambiguous connection between SJI 1400 grains and DFs in \ha\ or \calong. We minimized these effects in our study by using datasets from regions close to disk center where the field is more oriented along the line-of-sight. For further study of the spectral details of the bright grains, as function of time, one would ideally orient the \iris\ slit parallel to main axis of the DFs. No such datasets currently exist that include SST co-observations. 

We focussed mainly on the bright grains in SJI 1400 and their connection to DFs observed in \ha. We note that SJI 1330 images show the same grains with similar morphology, contrast, and dynamical properties as in SJI 1400. SJI 2976 however, does not show any clear sign of either grain like structures or DFs. While this filter is centered on the chromospheric \ion{Mg}{2}~k line, we attribute the absence of clear DF-associated signal to the relatively wide transmission profile (4~\AA) which allows for a non-negligible sub-chromospheric contribution. Despite the lower spatial resolution, the DFs can be identified in the EUV AIA channels as dark features against a bright background which we attribute to hydrogen and helium bound-free absorption. The presence of a bright rim at the DF top where we detect SJI 1400 and 1330 emission could not be unambiguously established in the AIA channels -- probably mostly due to the lower spatial resolution of AIA. While the EUV channels are mostly coronal, they all have (weak) contribution from transition region lines so one could expect EUV signal where we have clear TR signal in SJI 1400. We see little to no signs of the grains directly in the AIA 1600 channel which includes the transition region \ion{C}{4} line, likely because the broad passband \citep[$\sim$150~\AA, see ][]{boerner2012} is dominated by continuum emission, in particular by p-modes, and due to the small size of the grains relative to the AIA pixel size. While there are a few bright grains that might be associated with a similar signal in AIA 1600 it is not noticeable in the data unless one does running-difference movies. Even then the connection is extremely weak and does not jump out even with the benefit of comparing directly with the IRIS movies. The AIA 1600 and 1700 running difference movies are to a large degree similar but they show subtle differences in some locations at times, this may be due to transition region \ion{C}{4} emission in the AIA 1600. 

Our results support the conclusions of \cite{de_wijn_et_al_2006} that DFs sometimes reach transition region temperatures. Our results also confirm the findings of \cite{de-pontieu_et_al_2015} that magneto-acoustic shocks driven from below cause significant spectral broadening of the \silong\ lines. If we assume ionization equilibrium, the observations of \ion{S}{4} lines implies heating of the plasma in DFs to transition region temperatures. Type \footnotesize II\normalsize spicules \citep{de-pontieu_et_al_2007b} have also recently been shown to be heated to transition region temperatures, assuming ionization equilibrium \citep{pereira_et_al_2014, skogsrud2015, rouppe-van-der-voort_et_al_2015}. A recent study by \cite{olluri_et_al_2015} used state-of-the-art 3D magneto-hydrodynamics simulations and show that the \silong\ ion is out of ionization equilibrium in their simulated chromosphere and transition region. They find significant abundance of \ion{Si}{4} at lower temperatures than typically adopted for instantaneous ionization. This may lead to an overestimation of the temperature since \silong\ is used for temperature diagnostics. Further study is required to assess whether the \ion{Si}{4} emission we observe originates from plasma at transition region temperatures.  We observe possible heating toward the top of the DFs, while observations of type \footnotesize II\normalsize spicules suggest that they are heated along their entire length causing the entire structure to fade quickly in the cooler passbands. This indicates that there are at least two different heating processes at work. 

Based on the tight correlation we observe between chromospheric and transition region dynamics we conclude that a large fraction of transition region dynamics above active regions is caused by shock waves from below. However, the dark regions sometimes seen in the SJI 1400 slit-jaw images have no clear origin in our chromospheric and photospheric observations and are likely more affected by the overlaying corona. It remains an open question whether the waves impacting the transition region can propagate all the way into the corona \citep{de-pontieu_et_al_2005}.

Our observations and analysis show the complex nature of the transition region emission at the foot points of hot coronal loops. Previous interpretations have focused on interpreting this emission in terms of a classical transition region that is dominated by events and energy deposition from the overlying hot corona. The interpretation of intensity \citep[e.g. differential emission measure, ][]{pottasch1964}, Doppler shifts \citep[e.g. ``pervasive redshifts'', ][]{doschek1976} and line widths \citep[e.g. nanoflare heating, ][]{parker1988} of low transition region spectral lines has thus mostly focused on attempting to diagnose coronal conditions. Our observations suggest that this approach ignores the important, if not dominant, effects of the highly dynamic chromosphere on spectral lines that are formed in the transition region. The upper chromosphere is permeated with highly dynamic and violent events including shock waves and jets. Here we focused on the effects of so-called dynamic fibrils, jet-like features that appear to be driven by magneto-acoustic shock waves that propagate along the magnetic field in the low plasma $\beta$ environment of the plage upper chromosphere. Our results indicate that transition region emission associated with these ubiquitous events is responsible for a large fraction of the dynamic events seen in low transition region images such as the \iris\ SJI 1400 and SJI 1330 images. As these strong shocks impact the overlying transition region they cause short-lived blue shifts, excessive line broadening and intensity increases in the Si IV lines. Even after the initial impact of the shock, the dynamic evolution of dynamic fibrils continues to impact the low transition region emission, mostly at the top of the dynamic fibrils. It remains unclear from the current analysis how much of a role the returning plasma in the down-falling dynamic fibrils plays in the ``pervasive redshifts'' observed in the transition region. It also remains unclear what role the second, much faster, class of spicules, which are presumably driven by magnetic energy release, play in the general make-up of the transition region emission. Our earlier results suggest that, at the limb at least, these type \footnotesize II\normalsize spicules are often associated with down-flowing plasma at transition region temperatures, which suggests that they may in part be responsible for the pervasive redshifts. However, many questions remain that cannot be answered with the currently available datasets. Future coordinated datasets that optimize viewing angles, raster cadence and spatial coverage will be required to address how this mix of different types of spicules, flows and heating patterns combine to impact the emission in the low transition region.

\acknowledgements

IRIS is a NASA Small Explorer mission developed and operated by LMSAL with mission operations executed at NASA ARC and major contributions to downlink communications funded by the NSC (Norway). The Swedish 1--m Solar Telescope is operated on the island of La Palma by the Institute for Solar Physics (ISP) of Stockholm University in the Spanish Observatorio del Roque de los Muchachos of the Instituto de Astrof\' isica de Canarias. The research leading to these results has received funding from both the Research Council of Norway and the European Research Council under the European Union's Seventh Framework Programme (FP7/2007-2013) / ERC grant agreement nr. 291058. B.D.P. is supported by NASA contract NNG09FA40C (IRIS), and NASA grants NNX11AN98G and NNM12AB40P.

\bibliographystyle{apj}

\end{document}